\documentclass[11pt,a4paper]{article}
\pdfoutput=1
\usepackage[utf8]{inputenc}
\usepackage{arxiv}
\RequirePackage{fancybox}
\RequirePackage{amsmath}
\RequirePackage[]{graphicx}
\RequirePackage{amssymb}
\RequirePackage{amsfonts}
\usepackage[font=footnotesize,labelfont=bf]{caption}
\RequirePackage{natbib}
\graphicspath{{fig/}{../fig/}{./fig/}}
\AtBeginDocument{\DeclareGraphicsExtensions{.pdf,.PDF,.eps,.EPS,.png,.PNG,.tif,.TIF,.jpg,.JPG,.jpeg,.JPEG}}

\newcommand{\myv}[1]{\ensuremath{\mathbf{#1}}} 
\newcommand{\gv}[1]{\ensuremath{\mbox{\boldmath$ #1 $}}}  
\newcommand{\condexp}[3]{\mathcal{E}_{#1} \Big[ #2 \Big\vert #3 \Big]}
\newcommand{\ud}{\mathrm{d}}

\DeclareMathOperator*{\argmax}{arg\,max}

\begin{document}

\title{Model error covariance estimation in particle and ensemble Kalman filters using an online expectation-maximization algorithm}

\author{
Tadeo Javier Cocucci \\
FaCENA, Universidad Nacional del Nordeste\\
Corrientes, Corrientes, 3400, Argentina\\
\texttt{tadeojcocucci@gmail.com} \\
\And
Manuel Pulido \\
FaCENA, Universidad Nacional del Nordeste\\
Corrientes, Corrientes, 3400, Argentina\\
CONICET, Corrientes, Argentina\\
\And
Magdalena Lucini \\
FaCENA, Universidad Nacional del Nordeste\\
Corrientes, Corrientes, 3400, Argentina\\
CONICET, Corrientes, Argentina\\
\And
Pierre Tandeo\\
IMT Atlantique,\\
Lab-STICC, UMR CNRS 6285\\
F-29238, France\\
}

\maketitle

\begin{abstract}
The performance of ensemble-based data assimilation techniques that estimate
the state of a dynamical system from partial observations depends crucially on
the prescribed uncertainty of the model dynamics and of the
observations. These are not usually known and have to be inferred. Many
approaches have been proposed to tackle this problem, including  fully
Bayesian, likelihood maximization and innovation-based techniques. This work
focuses on maximization of the likelihood function via the
expectation-maximization (EM) algorithm to infer the model error covariance
combined with ensemble Kalman filters and particle
filters to estimate the state. The classical application of the EM algorithm
in a data assimilation context involves filtering and smoothing a fixed batch
of observations in order to complete a single iteration. This is an
inconvenience when using sequential filtering in high-dimensional
applications. Motivated by this, an adaptation of the algorithm
that can process observations and update the parameters on the
fly, with some underlying simplifications, is presented. The proposed
technique was evaluated and achieved good performance in experiments with the
Lorenz-63 and the 40-variable Lorenz-96 dynamical systems designed to
represent some common scenarios in data assimilation such as non-linearity,
chaoticity and model misspecification.
\end{abstract}

\section{Introduction}
Data assimilation techniques including variational data assimilation, ensemble
Kalman filters (EnKF) and particle filters estimate the state of a time-varying
process combining constraints on the dynamics of the system and some
observations of the process. Both the observation process and the dynamical
model of the state are prone to errors. Model error encompasses our incomplete
and possibly truncated knowledge of the dynamics including discretization
errors. On the other hand, observational error includes the inaccuracy of the
measurement instrument and the possible misrepresentation of the observational
operator that maps the state to the observations, also called representativity
error. Data assimilation relies on the specification of these errors to assess
the quality of the information provided by the forecast, given by the model
dynamics, and by the observations in order to provide an estimate of the state
that considers both sources of information. However, model and observational
errors are not usually available so they have to be estimated. A variety of
methods has been proposed to do this, including Bayesian approaches,
innovation-based methods and likelihood maximization techniques, among
others. A recent and detailed review of these methods can
be found in \citet{tandeo18}.

In particle or ensemble-based filtering techniques, the model error covariance
is related to the spread of the particles --ensemble members-- representing
the forecasted state. Using inadequate estimations of it may lead to a
misrepresentation of the forecast uncertainty which can produce filter
underperformance (when the spread is overdispersive), or in some cases,
its divergence (when the spread is underdispersive). Even if the true model
error covariance is used, it is well known (as shown in \citealp{houtekamer98})
that due to subsampling the ensemble has a tendency to collapse and possibly
lead to filter degeneracy. This  effect is often alleviated through a
multiplicative or additive covariance inflation factor
\citep{anderson99, miyoshi11}. Since the effective model error covariance
impacts on the ensemble spread, the problem of estimating these uncertainties
becomes important when using the EnKF.

In the case of particle filters, model error covariance impacts directly on
the uncertainty of each (independent) particle. Recently, some ideas have
emerged to improve the sampling of particle filters in high-dimensional spaces
guiding the particles toward regions of high likelihood. These approaches
include the implicit particle filter \citep{chorin09, atkins13}, the implicit
equal-weight particle filter \citep{zhu16}, the particle flow filter based on
tempering \citep{daum09} and the variational mapping particle filter
\citep{pulido_kernel_18}. An implicit assumption in all these filters is the
knowledge of the model error covariance. In other words, the knowledge of the
uncertainty associated to each particle. Therefore, these filters require in
practice a method to infer the model error covariance in complex systems.

There are several applications in which the estimation of the state of the
system is a big challenge because of the high-dimensionality \citep{snyder08}.
Estimation of model uncertainty is even more challenging in terms of
dimensionality. In several scenarios model error is represented through the
covariance matrix  $ \ensuremath{\mathbf{Q}} $ of a multivariate Gaussian distribution, so if the
dimension of the state is $ N_x $ then $\myv Q$ is a $N_x \times N_x$ matrix.
Hence, the estimation of the full covariance matrix $ \myv Q $ becomes
infeasible. To make this estimation problem achievable within
a data assimilation context, a commonly used approach is to parametrize the
matrix or assume it has a certain shape, such as band-diagonal. These
assumptions may be motivated physically, e.g. physical space localization of the
covariance \citep{hamill01, ueno10} or correlation functions in the sphere
\citep{gaspari99}.

In a similar fashion, the observational error covariance is often represented
by a $N_y \times N_y$ covariance matrix $\myv R$ of a multivariate Gaussian
distribution. Although $N_y$ is usually much smaller than $N_x$
parameterizations and fixed structure assumptions can be used to simplify the
problem of observational errror covariance estimation and to diminish
computational costs. For example, in \citet{stroud18}, a Matérn covariance
model is used to represent spatially correlated observation errors.

Some of the methods developed to estimate these error covariances focus on
finding the parameters that maximize the likelihood of the observations which
is usually computationally expensive to evaluate or approximate. In a seminal
work, \citet{shumway82} proposed to use the expectation-maximization (henceforth
EM) algorithm \citep{dempster77} in combination with the linear Kalman filter,
to produce a sequence of estimates that converges to the maximum likelihood
estimator. More recent applications of EM in EnKFs
\citep{tandeo15, dreano17, pulido_EM_18} use a fixed batch of observations that
are assimilated with a forward EnKF and a backwards ensemble Kalman Smoother in
order to complete one iteration of the EM algorithm. This batch EM approach was
extended by \citet{lucini19} for particle filters based on the innovation
likelihood to avoid the need of a particle smoother in the likelihood function.
The batch EM has a high-computational cost in high-dimensional spaces.
Furthermore,  the requirement to store and to reuse every observation within
the batch is inconvenient in the context of sequential filtering.

On the other hand, a variety of sequential or \textit{online} methods to
estimate error covariances has been developed. These produce updates ofthe
parameter estimations sequentially, at each assimilation cycle/filtering step,
so that there is no need to store all previous observations or to smooth
backwards through the whole batch of observations. An online version of the EM
algorithm has been proposed for sequential importance resampling (SIR) particle
filters in \citet{cappe09}, whereas \citet{andrieu03} also provided an online
version of EM for state-space models based on a split-data likelihood function
which combines pseudo-likelihood values computed with different mini-batches of
data. Based on these ideas, we propose here an efficient EM-based online
algorithm which can be coupled with both particle filters and EnKFs. Its
performance is assessed in experiments of interest in geophysical applications.

The rest of this work is outlined as follows. In Section
\ref{sec:theoretical_derivation}, we derive the proposed online method starting
from the batch EM algorithm. Section \ref{sec:experiments_and_results} describes
the experiments for which we evaluated the method's performance and examine the
results. In Section \ref{sec:conclusions}, we draw the conclusions and give
perspectives of possible future work. The appendices include some specification
of the dynamical systems (Appendix A) and the filtering techniques used in the
experiments (Appendix B).

\section{Theoretical derivation} \label{sec:theoretical_derivation}
We consider a hidden Markov model consisting in a dynamical model $\mathcal M$
for the hidden state variables $\myv x_k$ and observations of this state $\myv y_k$
at time $k$. The latter are represented through an observational function
$\mathcal H$ both with additive noise processes. Namely,
\begin{align}
\myv x_k &= \mathcal{M}(\myv x_{k-1}) + \gv \eta_k \label{Meq}, \\
\myv y_k &= \mathcal{H}(\myv x_k) + \gv \nu_k \label{Heq},
\end{align}
where the model and observational noise are realizations from
$\gv \eta_k \sim \mathcal N(\myv 0, \myv Q_k)$ and
$\gv \nu_k \sim \mathcal N(\myv 0, \myv R_k)$ respectively.

Data assimilation techniques involve estimating the density of the hidden state
given a set of observations, $p(\myv x_k|\myv y_{1:k})$, from Equations \ref{Meq}
and \ref{Heq} but rely on knowing the parameters of the state space model
including $\mathcal M$, $\mathcal H$, $\myv R_k$, $\myv Q_k$ and the initial state
density $p(\myv x_0)$.  The methodology to be described may estimate all of these
parameters, which we denote in general by $\gv \theta$. However, in this work,
the  dynamical model $\mathcal{M}$, observational operator $\mathcal{H}$ and
$p(\myv x_0)$ are assumed to be known, and we focus on the uncertainty
quantification of the hidden Markov model --the estimation of the error
covariances $\myv R_k$ and $\myv Q_k$ --, particularly the latter. The role of the
initial state density, $p(\myv x_0)$, is not expected to be important in
sequential assimilation apart from some initial spinup time.

\subsection{Batch EM algorithm}

The aim of the traditional batch EM algorithm applied to a hidden Markov model
is to iteratively find the statistical parameters $\gv \theta$ that maximize the
complete likelihood function $p(\myv y_{1:K} ; \gv \theta)$, given a batch of
observations distributed in a time interval,
$\myv y_{1:K}\triangleq \{\myv y_1,\dots,\myv y^{}_K\}$.  Since the realizations of
the state $\myv x_{0:K}$ are unknown, we write the likelihood function
in a  marginalized form,

\begin{equation}
  p(\myv y_{1:K} ; \gv \theta) =
  \int p(\myv {x}_{0:K}, \myv {y}_{1:K} ; \gv \theta) \ud \myv x_{0:K},
\end{equation}

The logarithm of this likelihood function may be reexpressed in terms of
an arbitrary density $q(\myv x_{0:K})$  \citep{neal98} with the sole constrain
that its support contains the support of the conditional density
$p(\myv x_{0:K}|\myv y_{1:K})$. Using the conditional density definition and that
$\int q(x_{0:K}) dx_{0:K} = 1$, the loglikelihood function can be written as,

\begin{align}
\log p(\myv y_{1:K} ; \gv \theta)  &= \mathcal D_{KL}(q(\myv x_{0:K}) \rVert p(\myv x_{0:K}|\myv y_{1:K}; \gv \theta)) + \int q(\myv x_{0:K}) \log  \frac{p_\theta(\myv x_{0:K},\myv y_{1:K} ; \gv \theta)}{q(\myv x_{0:K})} \ud \myv x_{0:K}, \nonumber \\
&\triangleq \mathcal D_{KL}(q\rVert p) + \mathcal{L}(q,\gv \theta), \label{evidence}
\end{align}
where the Kullback-Leibler divergence ($\mathcal D_{KL}(q|p) \triangleq \int
q(x) \log [q(x)/p(x)] \ud x$) is a non-negative function so that
$\mathcal{L}(q,\gv \theta)$ is a lower bound of the loglikelihood,
and is thus referred to as the evidence lower bound or ELBO
\citep{goodfellow16}. The upper bound of the ELBO is attained at
$q(\myv x_{0:K}) =  p(\myv x_{0:K}|\myv y_{1:K} ; \gv \theta)$ for which
$\mathcal D_{KL}(q|p)=0$. In this upper limit, the ELBO equals the
loglikelihood function.

Given an initial guess $\gv \theta_0$, the i-th iteration of the EM algorithm
encompasses
\begin{itemize}
\item {\em  Expectation step.} Maximize the ELBO as a function of $q$ given a
set of parameters $\gv \theta_{i-1}$, i.e.
$q_{i-1}= \argmax\limits_q  \mathcal L(q,\gv \theta_{i-1})$.
The resulting optimal density is
$q_{i-1}(\myv x_{0:K})=p(\myv x_{0:K}|\myv y_{1:K};\gv \theta_{i-1})$.
\item {\em  Maximization step.} Maximize the ELBO w.r.t. $\gv \theta$ given a
fixed $q_{i-1}$, i.e.
$\gv \theta_{i}= \argmax\limits_{\gv \theta} \mathcal L( q_{i-1}, \gv \theta )$.
\end{itemize}
Since the methodology guarantees that $\mathcal L(q,\gv \theta)$ increases at
each EM iteration (or remains unchanged if a maximum has been reached), while
$\mathcal D_{KL}(q_i(\myv x_{0:K}) \rVert p(\myv x_{0:K}|\myv y_{1:K}; \gv \theta))$
decreases (or remains unchanged), the algorithm converges to a set of parameters
that  (locally) maximizes the loglikelihood \citep{wu83}. In other words, the
approximation function $q(\myv x_{0:K})$ during the convergence  gets closer to
the posterior density given the optimal parameters as the ELBO increases and
therefore the Kullback-Leibler divergence between these two probability density
functions decreases.

A practical aspect of the algorithm is that the two steps do not require to
reach the maximum of the ELBO to guarantee convergence. As long as we find a
$\gv \theta_i$ that increases the ELBO in the maximization step (this family
of methods are called generalized EM algorithms) and a function  $q_{i-1}$ that
increases the ELBO in the expectation step (called as incremental EM
algorithms), the convergence is guaranteed.

\subsection{The batch EM algorithm in a hidden Markov model}

A central property of the hidden Markov model is that the joint density can be
expressed as a product. Using the Markov property of the state, and that
observations only depend on the current state, the joint probability density
in a time interval $0{:}K$ (e.g. \citealp{pulido_EM_18}) results in
\begin{align}
p(\myv x_{0:K}, \myv y_{1:K} ; \gv \theta) &= p(\myv x_0 ; \gv \theta) \prod_{k=1}^{K} p(\myv x_{k} | \myv x_{k-1} ; \gv \theta) \prod_{k=1}^{K} p(\myv y_{k} | \myv x_{k} ; \gv \theta) \nonumber \\
 &=  p(\myv x_0 ; \gv \theta)   \prod_{k=1}^{K} p(\myv x_{k}, \myv y_{k} | \myv x_{k-1}; \gv \theta). \label{MarkovJoint}
\end{align}
Using the product form of the joint density, in Equation \ref{MarkovJoint},
the resulting ELBO function for hidden Markov models is
\begin{align}
  \mathcal L(p(\myv x_{0:K}|\myv y_{1:K} ; \gv \theta_{i-1}) , \gv \theta) & = \sum_{k=1}^K \int p(\myv x_{k-1:k} |\myv y_{1:K}; \gv \theta_{i-1})  \log p(\myv x_k, \myv y_k |\myv x_{k-1}; \gv \theta) \ud \myv x_{k-1:k} + C, \nonumber \\
& \triangleq \sum_{k=1}^K\condexp{}{\log p(\myv {x}_{k}, \myv {y}_{k} | \myv {x}_{k-1})} {\myv y_{1:K}; \gv \theta_{i-1}} + C, \label{eq:G}
\end{align}
where all the constant terms w.r.t. $\gv\theta$ are included in $C$ and dropped
from $\mathcal L$ in what follows. Furthermore, note that
$\condexp{}{f (\myv x_{k-1}, \myv x_k)} {\myv y_{1:K} ; \gv \theta} = \int f(\myv x_{k-1}, \myv x_k) p(\myv x_{0:K}| \myv y_{1:K} ; \gv \theta) \ud \myv x_{k-1:k} = \int f(\myv x_{k-1}, \myv x_k) p(\myv x_{k-1}, \myv x_k|\myv y_{1:K} ; \gv \theta) \ud \myv x_{k-1:k}$.

In the general case, the parameters that maximize the ELBO in a hidden Markov
model in the maximization step, need to be determined numerically.
However, if $p(\myv {x}_{k}, \myv {y}_{k} | \myv {x}_{k-1})$ belongs to an exponential
family (condition satisfied when both observational and model errors belong to
the exponential family), it is possible to derive an analytical expression for
the parameters that maximize the ELBO (Equation \ref{eq:G}). This is a rather
general parametric density family assumption which includes the usual Gaussian
model and observational errors, to be specified below. The joint density of
state and observation  given the previous state required in Equation
\ref{MarkovJoint} is in that case expressed as
\begin{equation}\label{eq:pdfexpfamily}
p (\myv x_{k}, \myv y_{k} | \myv x_{k-1} ; \gv \theta) = h(\myv x_{k}, \myv y_{k}) \exp \left[\psi (\gv \theta) \cdot s(\myv x_{k-1}, \myv x_{k}, \myv y_{k}) -A(\gv \theta)\right],
\end{equation}
where $s(\myv x_{k-1}, \myv x_{k}, \myv y_{k})$ is the natural sufficient statistic,
$\psi(\gv \theta)$ is called the natural parameter and $h$ and $A$ are functions.

The gradient of the ELBO w.r.t.  $\gv\theta$ by introducing Equation
\ref{eq:pdfexpfamily} into \ref{eq:G} is
\begin{align}
\nabla_\theta \mathcal L (p(\myv x_{0:K}|\myv y_{1:K} ; \gv \theta_{i-1}) , \gv \theta)  =\nabla_\theta \psi(\gv\theta) \cdot \sum_{k=1}^K \condexp{}{s(\myv x_{k-1}, \myv x_{k}, \myv y_{k})} {\myv y_{1:K}; \gv \theta_{i-1}} - K \nabla_\theta A(\gv \theta). \label{gradL}
\end{align}

Assuming the expression in Equation \ref{gradL} has one root, corresponding to
the maximum of $\mathcal L$, the resulting E-step of the i-th iteration becomes
\begin{align}
S_{i-1} = \frac{1}{K} \sum_{k=1}^{K} \condexp{}{s(\myv x_{k-1}, \myv x_{k}, \myv y_{k})} {\myv y_{1:K}; \gv\theta_{i-1}}, \label{SK}
\end{align}

while the M-step is
\begin{align}
  \gv\theta_{i} = \widehat{\gv\theta}\left(S_{i-1}\right) \label{hattheta},
\end{align}
where we define $\widehat{\gv\theta}(S_{i-1})$ as the solution for $\gv \theta$
of the equation
$\nabla_\theta \psi(\gv\theta) \cdot S_{i-1} - K \nabla_\theta A(\gv \theta) = 0$.

In the case that both the observational and the transition densities are Gaussian
\begin{align*}
p(\myv {x}_{k} | \myv {x}_{k-1}) &= \mathcal{N}(\mathcal{M}(\myv {x}_{k-1}), \myv Q), \\
p(\myv {y}_{k} | \myv {x}_{k}) &= \mathcal{N}(\mathcal{H}(\myv {x}_{k}), \myv R),
\end{align*}
and the parameters to estimate are their respective covariances
$\gv \theta = \{\myv Q, \myv R\}$ which are assumed constant in time,
the quantity $S_{i-1}$ can be expressed as a tuple $(S_{i-1}^Q, S_{i-1}^R)$.
Using Equations \ref{gradL} and \ref{SK} we get the corresponding expressions,
\begin{align} \label{eq:update_batch}
S_{i-1}^{Q} &= \frac{1}{K} \sum_{k=1}^{K} \condexp{}{ (\myv x_k - \mathcal{M}(\myv x_{k-1}))(\myv x_k - \mathcal{M}(\myv x_{k-1}))^\top}{\myv y_{1:K};\gv \theta_{i-1}}, \\
S_{i-1}^{R} &= \frac{1}{K} \sum_{k=1}^{K} \condexp{}{(\myv y_k - \mathcal{H}(\myv x_{k}))(\myv y_k - \mathcal{H}(\myv x_{k}))^\top}{\myv y_{1:K};\gv \theta_{i-1}}.\label{eq:update_Rbatch}
\end{align}

In this particlar case, because of the Gaussian assumption, the maximization
with respect to the parameters in Equation \ref{hattheta} results in
$(\myv Q_{i}, \myv R_{i}) = \widehat{\gv\theta}(S_{i-1}) = S_{i-1} = (S_{i-1}^Q, S_{i-1}^R)$,
which amounts to say that $\widehat{\gv\theta}$ is the identity function.

The update formulas in Equations \ref{eq:update_batch} and
\ref{eq:update_Rbatch} are the main calculations used in the implementations of
batch EM such as \citet{tandeo15, dreano17, pulido_EM_18}.
These expressions involve computing the smoothed densities
$p(\myv x_{k}|\myv y_{1:K})$ for every $k$ at each EM iteration through Kalman
smoothing techniques in the Gaussian linear hidden Markov model. In other words,
this requires to process the whole batch of observations in each iteration of
the EM algorithm. Furthermore, if new observations are added, the whole process
needs to be redone. The batch EM is a robust way to estimate the global
structure of $\myv Q$ and $\myv R$. However, since it can be very costly and
unsuitable in the context of observations gathered sequentially in time, we aim
to develop an ``online'' approach for high-dimensional systems based on the EM
algorithm that enables to update the parameters \textit{at each} assimilation
cycle. The goal of doing this is to avoid to store all of the observations and
to reduce substantially the computational cost, since each observation would be
processed only once. Besides, online techniques may track time-varying error
covariances which is important for geophysical applications.

\subsection{The online EM algorithm}

The major drawback of batch EM is that for each new observation, and a given
value of the parameters $\gv \theta$, the ELBO has to be recomputed using the
whole sequence of observations from $1$ to $K$. Our goal is to update the
parameter with each new observation. Thus, in what follows the EM iteration
index is dropped because, in this sequential context, it coincides with $K$,
the index of the last observation being processed. With this in mind, we will
derive here an approximated recursive form for the required quantity in Equation
\ref{SK}. We begin by writing it as:
\begin{align}
S_K &= \frac{1}{K} \left( \sum_{k=1}^{K-1} \int s(\myv x_{k-1}, \myv x_{k}, \myv y_{k})
p(\myv x_{k-1}, \myv x_k | \myv y_{1:K} ; \gv\theta_K) d\myv x_{k-1:k} + \right. \nonumber \\
 & \left. \int s(\myv x_{K-1}, \myv x_{K}, \myv y_{K})
  p(\myv x_{K-1}, \myv x_K | \myv y_{1:K} ; \gv\theta_K) d\myv x_{K-1:K} \right). \label{sum_memory}
\end{align}

We can recognize that the first $K-1$ terms in Equation \ref{sum_memory}
correspond to $(K-1)S_{K-1}$ but incorporating information of the newly
available observation $\myv y^{}_K$ so that the posterior density corresponds to
smoothing. We make the approximation that $\myv y^{}_K$ does not significantly
influence the previous state estimates but only provides information to the last
term, which corresponds to the sufficient statistics at $K$. This results in:

\begin{align}
\tilde{S}_{K} &= (1-\gamma_K) \tilde{S}_{K-1}
+ \gamma_K \int s(\myv x_{K-1}, \myv x_{K}, \myv y_{K})
p(\myv x_{K-1}, \myv x_K | \myv y_{1:K}; \gv\theta_{K}) d\myv x_{K-1:K} \\
&= \left( 1-\gamma_K \right) \tilde{S}_{K-1} +
\gamma_K \mathcal{E}_{} \left[ s(\myv x_{K-1}, \myv x_{K}, \myv y_{K}) | \myv y_{1:K}; \gv\theta_{K} \right], \label{eq:suf_stats_update}
\end{align}
where we introduced $\gamma_k \in (0, 1)$ as a step-size (instead of $1/K$),
based on stochastic approximation techniques \citep{legland97}. This parameter
controls the memory of the statistics, determining the importance of the old
statistics relative to the contribution of the current observation. The
initialization parameter $\tilde{S}_0$ has to be provided as a first
guess. In  the Gaussian case, it coincides with the first guess for the
parameter $\gv\theta_0$.

There is some resemblance with the technique proposed by
\citet{cappe09} where, in order to keep track of the statistics, an
auxiliary function needs to be kept updated. This auxiliary function relates
to a recursive form of smoothing and can be used to propagate the information
of the latest observation to the previous estimate of the statistic. In our
algorithm, $\gamma_k$ controls the influence of previous estimate on the
current one, so under the proposed assumption, we bypass the need of using
this auxiliary function. The auxiliary function in \citet{cappe09}
makes an exact EM update. On the other hand, the algorithm here proposed,
is an approximation to account for just a single smoothing step. A stronger
assumption was taken in \citet{lucini19}, they proposed an EM batch algorithm
which only uses the filtering posterior density without accounting the effect
of subsequent observations during the parameter estimation.

To compute the expectation in Equation \ref{eq:suf_stats_update}, we propose two
related approaches: importance sampling integration implemented in such a way
that no smoothing is required (suitable for particle filters) and direct Monte
Carlo integration using a sample of the required joint smoothing density
$p(\myv x_{k-1:k} | \myv y_{1:k})$. These two methods are further explained in
the following subsections.

In the batch EM, the parameters are fixed during the whole batch of
observations and are updated in each EM iteration. On the other hand, the
parameters are updated in this online EM version at each observation time when
information on the current observations is available. On the other hand, in this
online version of the EM algorithm the parameters are updated when a new
observation is available, that is, at each observation time. Thus, while batch
EM assumes that parameters are fixed, in the present scheme we assume parameters
change slowly as in hierarchical Bayesian approaches
\citep{scheffler19}.

\subsubsection{Importance sampling}
\label{sec:importance_sampling}

In order to compute the expectation in Equation \ref{eq:suf_stats_update}, we
write it in integral form:
\begin{align}
\mathcal{E}_{}
\left[ s(\myv x_{K-1}, \myv x_{K}, \myv y_{K}) | \myv y_{1:K}; \gv\theta_{K} \right]
= \int s(\myv x_{K-1}, \myv x_{K}, \myv y_{K})
p(\myv x_{K-1:K} | \myv y_{1:K}; \gv\theta_{K}) d\myv x_{K-1:K} \label{expectation_s}
\end{align}
To conduct the integration in Equation \ref{expectation_s}, Bayes rule is used
to rewrite the density
\begin{align}
p(\myv x_{K-1:K} | \myv y_{1:K})
&= p(\myv x_{K-1} | \myv x_{K},  \myv y_{1:K-1})
p(\myv x_{K} | \myv y_{1:K}) \nonumber \\
&= \frac{p(\myv x_{K} | \myv x_{K-1}) p(\myv x_{K-1} | \myv y_{1:K-1})}
{p(\myv x_{K} | \myv y_{1:K-1})}
p(\myv x_{K} | \myv y_{1:K}) \nonumber\\
&= p(\myv x_{K} | \myv x_{K-1}) p(\myv x_{K-1} | \myv y_{1:K-1})
\frac{p(\myv y_{K} | \myv x_{K})} {p(\myv y_{K} | \myv y_{1:K-1})}. \label{denskk}
\end{align}
where the parameter $\gv \theta_{K}$ has been dropped from the density
expression for the sake of simplicity. Equation \ref{denskk} suggests that
instead of sampling from $p(\myv x_{K-1:K} | \myv y_{1:K} ; \gv\theta_K)$, we can
sample from
$p(\myv x_{K} | \myv x_{K-1} ; \gv\theta_K) p(\myv x_{K-1} | \myv y_{1:K-1} ; \gv\theta_K)$
and each particle should be weighted with the corresponding unnormalized
importance weight proportional to $p(\myv y_{K} | \myv x_{K} ; \gv\theta_K)$.
Finally, they should be normalized to sum one. This procedure is similar to the
sampling performed in the bootstrap filter \citep{gordon93, arulampalam02} in
which the particles from the marginalized posterior density at a previous time
$p(\myv x_{K-1} | \myv y_{1:K-1} ; \gv\theta_K)$ evolve forward through the
transition density, $p(\myv x_{K} | \myv x_{K-1} ; \gv\theta_K)$.
However, here we conduct a \textit{single} time step with the only purpose of
approximating the expectation (to estimate the parameters) so that no resampling
is required. We assume that a particle representation of the filtering posterior
density, $p(\myv x_k|\myv y_{1:k} ; \gv\theta_k)$, is available, which is the case
for sequential Monte Carlo techniques and EnKFs. In this work, we use the
perturbed observation EnKF \citep{burgers98} and the variational mapping
particle filter \citep{pulido_kernel_18} to infer the marginalized posterior
density. The sampling procedure to evaluate Equation \ref{expectation_s} is
independent of the filtering algorithm and is thus expected to be suitable for
any variant of the ensemble Kalman or particle filters.

In this way,  we can sample at $K-1$ using the filtering density  and then
obtain a sample of the forecast at time $K$ through the transition density. The
sampling points are denoted as
\begin{align*}
\{\myv x_{K-1}^{a(j)} \}_{j=1}^{N_p} \sim p(\myv x_{K-1} | \myv y_{1:K-1};\gv\theta_{K}),
\end{align*}
and for each $j$
\begin{align*}
\{\myv x_{K}^{f(j,l)} \}_{l=1}^{M_p} &\sim p(\myv x_{K} | \myv x_{K-1}^{a(j)};\gv\theta_{K} ),
\end{align*}
where the superscripts $a$ and $f$ stand for analysis and forecast. $N_p$ and
$M_p$ are the number of particles used to approximate each integral. The number
of particles provided by the filter is $N_p$ and for each of these a sample of
size $M_p$ is drawn using the transition density. Therefore, we can approximate
Equation \ref{expectation_s} via Monte Carlo integration as a double sum:
\begin{align}
\mathcal{E}_{}
\left[ s(\myv x_{K-1}, \myv x_{K}, \myv y_{K}) | \myv y_{1:K}; \gv\theta_{K} \right]
\approx \sum_{j=1}^{N_p} \sum_{l=1}^{M_p}
w_{j,l} \, s(\myv x_{K-1}^{a(j)}, \myv x_{K}^{f(j, l)}, \myv y_{K}), \label{doblesum}
\end{align}
where the importance weights $w_{j, l}$ are the normalized counterparts of the
unnormalized weights,
\begin{align*}
\bar{w}_{j,l} = p(\myv y_{K} | \myv x_{K}^{f(j,l)}).
\end{align*}
In cases in which the number of particles is large, $M_p$ could be chosen equal
to 1 producing from each particle at $\myv x_{K-1}^{a(j)}$ a forecast with a
single realization of model error. In this case a single sum remains in
Equation \ref{doblesum}.

\subsubsection{One-step smoother}
\label{sec:one_step_smoother}

An alternative to importance sampling is to directly sample from
$p(\myv x_{K-1:K} | \myv y_{1:K})$ assuming parametric statistics. In order to do
this, we need to compute or approximate the smoothing density
$p(\myv x_{K-1} | \myv y_{1:K})$. In the context of filtering with an EnKF, this can
be done using the ensemble variant of the RTS Kalman smoother (EnKS)
\citep{rauch65, cosme12}. This is a backwards algorithm that is usually applied
to get the smoothing densities for the state at all times, but here we are
interested in one of them, namely the smoothing density at time $K-1$, thus we
only need to do one step backwards of the algorithm. On the other hand, the
density of the last state conditioned on all observations is the filtering
density that is provided by the EnKF. Hence, if we have
\begin{align*}
\{ \myv x_{K-1}^{s(j)} \}_{j=1}^{N_p} &\sim p(\myv x_{K-1} | \myv y_{1:K};\gv\theta_{K}), \\
\myv x_{K}^{a(j)}  &\sim p(\myv x_{K} |\myv y_{1:K};\gv\theta_{K} ),
\end{align*}
we can approximate the expectation in Equation \ref{expectation_s} using Monte
Carlo integration as
\begin{align*}
\mathcal{E}_{}
\left[ s(\myv x_{K-1}, \myv x_{K}, \myv y_{K}) | \myv y_{1:K};\gv\theta_{K} \right]
\approx \frac{1}{N_p} \sum_{j=1}^{N_p}
s(\myv x_{K-1}^{s(j)}, \myv x_{K}^{a(j)}, \myv y_{K}).
\end{align*}

We generate the EnKS smoothed particles $ \myv x_{K-1}^{s(j)} $ following
\citet{cosme12} through the equations
\begin{align*}
\myv K_{K-1}^s &= \myv S_{K-1}^a [(\myv S_K^f)^T \myv S_K^f]^{-1} (\myv S_K^f)^T, \\
\myv x_{K-1}^{s(j)} &= \myv x_{K-1}^{a(j)} +
	\myv K_{K-1}^s (\myv x_{K}^{s(j)} - \myv x_{K}^{f(j)}),
\end{align*}
where $\myv S_{K-1}^a$ and $\myv S_{K}^f$ consists of particles of the analysis at
time $K-1$ and of the forecast particles at time $K$ respectively, each particle
centered at their mean, i.e. $\myv x_{K}^{a(j)}- \overline{\myv x}_{K}^{a}$ and
$\myv x_{K}^{f(j)}- \overline{\myv x}_{K}^{f}$ , and set as columns. Also, note that
the smoothing distribution of the last state $\myv x_K$ coincides with the
filtering distribution, $\myv x_{K}^{s(j)}=\myv x_{K}^{a(j)}$.

This implementation of an online EM algorithm for the Gaussian case based on the
EnKF has some resemblance to the technique proposed in \citet{berry13}. That
work also proposes an exponential moving average for the parameters. However,
the updates of the parameters  are derived from analyzing the cross-correlation,
using different time lags, of the forecast based innovations
$\gv \epsilon_k = \myv y_k - \myv H \myv x_k^f$ (where $\myv H$ is the linealization of
the observational operator $\mathcal{H}$) which can be related to the matrices
$\myv Q$ and $\myv R$ once they have reached a stationary regime. On the other hand,
our approach is based on likelihood maximization. At evaluation of the
sufficient statistics, it makes use of filtered and smoothed estimates of the
state.  The expectation on Equation \ref{eq:suf_stats_update} is conditioned on
all available observations including the observation at time $K$.

\section{Experiments and results} \label{sec:experiments_and_results}

A series of twin experiments was conducted using a known model error covariance,
referred as true model error covariance, to evaluate the convergence of the
estimator. The chaotic dynamical systems used in the experiments are the
Lorenz-63, the one-scale Lorenz-96 and the two-scale Lorenz-96 systems for which
equations, parametrizations and implementation details are given in Appendix
\ref{app-A}. The proposed EM algorithm was implemented with two filters, an
ensemble Kalman filter that uses perturbed observations \citep{burgers98} and a
particle filter based on gradient flows, the variational mapping particle filter
(VMPF, \citealp{pulido_kernel_18}). Details of the two filters are given in
Appendix \ref{app-B}. For the proposed importance sampling methodology (Section
\ref{sec:importance_sampling}), we conducted a set of experiments using the
variational mapping particle filter (referred to as IS-VMPF) and another one
using the ensemble Kalman Filter (IS-EnKF). The set of experiments with the
one-step smoother and the EnKF is referred to as OSS-EnKF.

Details of the experiments performed to evaluate the methodology are: the
assimilation cycle length, time between observations, is $\Delta t = 0.05$
time units in every experiment. This corresponds to 5 integration time steps
for the Lorenz-63 (for which the integration time step is $\delta t = 0.01$) and
50 integration time steps for the Lorenz-96 and the two-scale Lorenz-96 system
($\delta t = 0.001$). We used $N_p=50$ particles for the filtering techniques
in experiments with the Lorenz-63 and the 8 variables Lorenz-96, and $N_p = 100$
for the experiment with the 40 variables Lorenz-96 system. For the
importance-sampling-based techniques, we used $M_p=20$ as the number of samples
to compute the Monte Carlo integration in all experiments except for the 40
variables Lorenz-96 for which we used $M_p = 100$. We use the notation
$\myv I_n$ to denote the $n \times n$ identity matrix, $N_x$ the state dimension
and $N_y$ the dimension of the observation space. We chose step sizes of the
form $\gamma_k = k^{-\alpha}$ where $k$ indexes the cycle and
$\alpha \in (0.5, 1)$ is a tunable parameter. With this scheme,
more weight is given to the current estimates in the initial iterations and
more weight is given to previous estimates in later cycles when the parameter
is expected to be already better estimated. The parameter $\alpha$ controls
how is the weight distributed between the current and previous estimations.
We chose $\alpha = 0.6$ for all experiments except stated otherwise. This choice
is based on manual tuning experiments which evaluate the performance for
different values of $\alpha$.

\subsection{Lorenz-63 system}
\label{sec:lor63}
The observations are generated from the 3-dimensional Lorenz-63 dynamical system
with Gaussian model and observational uncertainties using diagonal covariance
matrices $\myv Q = \sigma_Q^2 \myv I_{N_x}$ and $\myv R = \sigma_R^2 \myv I_{N_y}$ where
$\sigma_Q^2 = 0.3$ and $\sigma_R^2 = 0.5$. We estimate $\myv Q$ and assume $\myv R$
is known. We assume the full state is observed which implies that $N_y = N_x$.
The uncertainty in the estimator is empirically evaluated by repeating the
experiment 30 times with different initial samples and observations.

\begin{figure}[h]
\captionsetup{width=0.5\textwidth}
\centering
\includegraphics[scale=0.5]{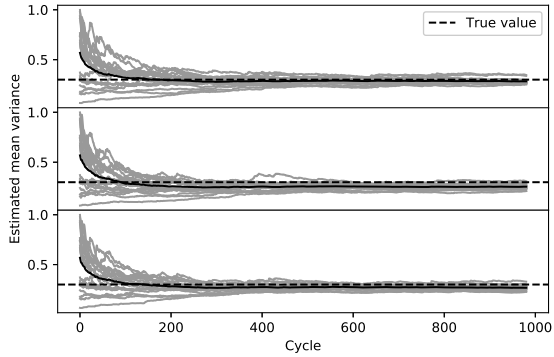}
\caption{Estimated mean model error variance with different first guesses for
the three different variations of the estimation algorithm,
(a): IS-EnKF, (b): OSS-EnKF, (c): IS-VMPF}
\label{fig:lor63_seeds}
\end{figure}

Figure \ref{fig:lor63_seeds}  shows the estimations of the different
realizations of the  Lorenz-63 experiment for the IS-EnKF,  OSS-EnKF and IS-VMPF
implementations. Estimations for different initial guesses converge
consistently. The estimated mean value is shown with a black line and a slight
underestimation of the true value can be detected. The realizations show the
convergence of the estimation's uncertainty which occurs after about 500 cycles.
This uncertainty also includes sampling noise due to the particle representation
of the state distribution.

\begin{figure}[h]
\captionsetup{width=0.5\textwidth}
\centering
\includegraphics[scale=0.5]{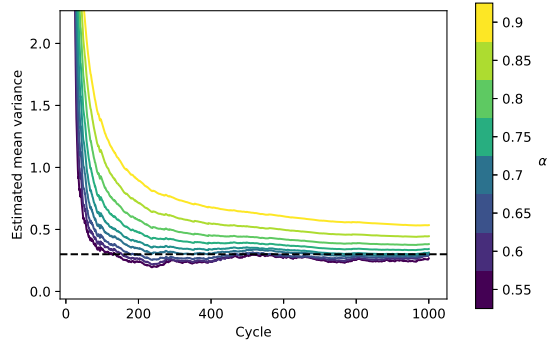}
\caption{Estimated mean model error variance for different values of
$\alpha$ in the step size sequences $\gamma_k = k^{-\alpha}$.}
\label{fig:lor63_learning_rates}
\end{figure}

Another experiment was conducted to evaluate the influence of the step size,
$\gamma_k$, on the convergence. The experiment was repeated with the same
observations but the step size changing according to $\gamma_k = k^{-\alpha}$
\citep{legland97} for different values of $\alpha$. The impact of different
choices for the step size on the estimated model error is shown in Fig.
\ref{fig:lor63_learning_rates}. Larger $\alpha$ can be interpreted as more
memory in the previous estimates, meaning that in
Equation \ref{eq:suf_stats_update}, more weight is given to previous model error
estimations leading to estimates with less noise. However, they stabilize
prematurely overestimating the true value. For smaller step sizes, the resulting
estimations are closer to the true value but are noisier. Only results for the
IS-VMPF implementation are shown, but results are analogous for IS-EnKF and OSS-EnKF.

\subsection{Lorenz-96 system}
An experiment with the  Lorenz-96 model was conducted in which the observations
are generated using a true model error covariance matrix with spatial
correlations. We consider positive model error covariance between neighbouring
variables (variables next to each other, considering a periodic domain) and null
covariance between non-neighbouring ones. The model error covariance between
each pair of neighbouring variables is set to $0.09$ and the model error
variance of each variable is set to $0.3$. This model error variance is
representative of a two-scale  Lorenz-96 model when the small-scale is ignored
\citep{pulido17, pulido_EM_18}. The observational error covariance matrix,
$\myv R = 0.5 \myv I_{N_y}$, is assumed known and only $ \myv Q$ is estimated.
The dynamical system used is the 8 variables Lorenz-96 model and the full state
is observed.

\begin{figure}[h]
\captionsetup{width=0.5\textwidth}
\centering
\includegraphics[scale=0.5]{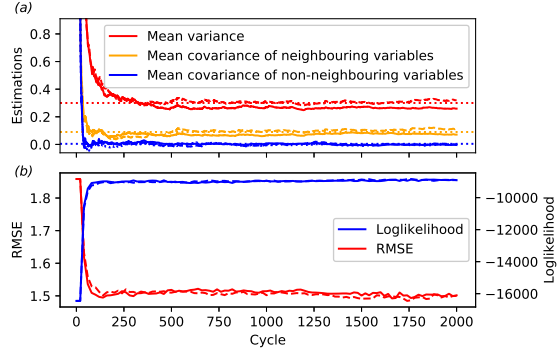}
\caption{(a): Estimated mean variance (red), mean covariance of neighbouring
variables (orange), and mean covariance of non-neighbouring variables (blue)
as a function of the assimilation cycle for the  OSS-EnKF experiment (solid
lines) and the  IS-VMPF experiment (dashed lines). True values are shown with
dotted lines. (b): Total RMSE (red line) and loglikelihood function (blue line)
computed for all the cycles using the corresponding estimated model error.}
\label{fig:lor96_covs}
\end{figure}

Figure \ref{fig:lor96_covs} (a) shows  the mean of the diagonal, the mean of
covariances of neighbouring variables and the mean of covariances of
non-neighbouring variables for the OSS-EnKF and IS-VMPF experiments. The mean
values are for the only purpose of presentation but note that each matrix entry
value is estimated. The major structure of the matrix is correctly recovered.
Results for IS-EnKF are not shown but similar results were obtained.

For a meaningful evaluation of the convergence of the method, the full set of
observations, $\myv y_{1:K}$,  needs to be considered in the likelihood funtion,
i.e. the complete likelihood function. However note that the parameters change
in the proposed online method after each observation is assimilated. To find an
approximation of the complete likelihood function, we computed state estimates
with a complete pass of an EnKF through all the observations, $\myv y_{1:K}$,
using a model error covariance that is fixed in the whole time interval $1{:}K$.
With the produced state estimates, we computed an approximation of the
loglikelihood. This is conducted for each $\myv Q_k$ (estimation of $\myv Q$ at the
k-th cycle) of the error covariance. In a similar way, the RMSE with respect to
the true state is determined. In Fig. \ref{fig:lor96_covs} (b), a rapid increase
of the loglikelihood and an consistent decrease of RMSE is found. Both
quantities then stabilize. This behavior, however, does not guarantee that a
global maximum has been reached but that estimations increase the loglikelihood
and improve RMSE performance quickly in the first iterations. The RMSE appears
to show some improvement even for very long cycles.

\begin{figure}[h]
\captionsetup{width=0.5\textwidth}
\centering
\includegraphics[scale=0.5]{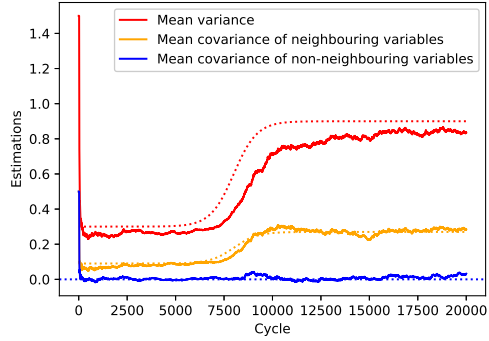}
\caption{Estimated mean variance (red), mean covariance of neighbouring variables
(orange), and mean covariance of non-neighbouring variables (blue) for the
OSS-EnKF experiment (solid lines). The corresponding true time varying values
are shown with dotted lines.}
\label{fig:lor96_covs_varyingQ}
\end{figure}

The Lorenz-96 experiment is repeated but using a true model error covariance
which slowly varies in time with a sigmoid behavior in every non-zero entry
value. Figure \ref{fig:lor96_covs_varyingQ} shows the estimated $\myv Q$ values.
The covariance structure of the matrix is identified and the estimations respond
to the slow variation in time of the true parameter. While the matrix increases
its values on the non-zero band, the estimations which correspond to values
outside this band slightly increase following the behavior of the rest of the
matrix. This is in a way compensated by the underestimation of the main
diagonal. We obtained similar results for all three implementations but only
show them for OSS-EnKF to avoid a cluttered figure. The step-sizes $\gamma_k$
control the sensitivity of the estimations to past estimations so we expect that
the ability of the algorithm to track changes in the error covariances is
related on how $\gamma_k$ is chosen. In particular we expect that if we choose
$\gamma_k$ in such a way that it gives more weight to previous estimations the
algorithm will fail to track fast changes in the error covariances. However we
do not study this behavior further since it is out of the scope if this work.

In order to assess the performance of the algorithm in a higher-dimensional
state space, we conducted an experiment using the 40-variables Lorenz-96 model.
Observations of the full state are available every assimilation cycle. We take
$\myv Q = \sigma_Q^2 \myv I_{N_x}$ and $\myv R = \sigma_R^2 \myv I_{N_y}$ with
$\sigma_Q^2=0.3$ and $\sigma_R^2=0.5$.

\begin{figure}[h]
\captionsetup{width=0.5\textwidth}
\centering
\includegraphics[scale=0.5]{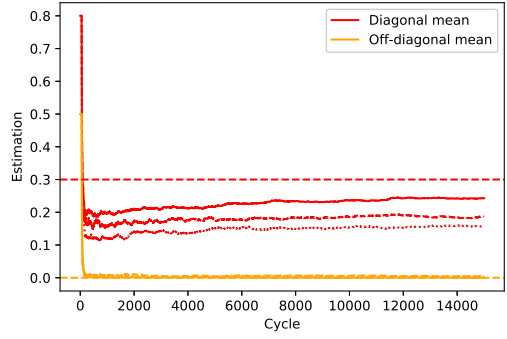}
\caption{Convergence results for estimations of $\myv Q$ in the 40-variable
Lorenz-96 system for the three different implementations of the proposed
algorithm, OSS-EnKF (continuous line), IS-VMPF (dashed line) and  IS-EnKF
(dotted line).}
\label{fig:lor96_results}
\end{figure}

Figure \ref{fig:lor96_results} shows the estimations of the mean of the diagonal
and off-diagonal values of the model error covariance as a function of the
assimilation cycle. IS-VMPF and OSS-EnKF produce reasonable results, although
with a tendency to underestimate. IS-EnKF exhibits the largest underestimation
of model error variance. This could in part be explained by the tendency of the
EnKF to collapse and because in importance sampling the
computation of weights in high dimensions often gives very small
unrepresentative values. This underestimation is found for relatively small
values of observational error variance, experiments with larger observational
errors tend to give better estimates, with slight overestimations as is shown
in the next section. The underestimation is not severe enough to induce filter
divergence. Additionaly, some slight improvement is found in the three
experiments for latter iterations after the fast changes in the first
iterations. We note that no inflation factor in the EnKF is used for these
experiments. The values outside the diagonal, which correspond to zero, are
correctly estimated with every technique.

\subsection{Impact and estimation of observation error covariance matrix}
The estimation of $\myv Q$ and $\myv R$ are closely related \citep{tandeo18}
and often $\myv Q$ compensates for a badly specified $\myv R$. In order to assess
the influence of having different magnitudes of the observational error variance
in the estimation of $\myv Q$, we conducted two experiments:

- \textit{(i)} For an 8-variables Lorenz-96 system, we estimate
$\myv Q = 0.3 \myv I_{N_x}$ using different values for the observational error
variance $\myv R = \sigma_R^2 \myv I_{N_y}$ and assuming its true value is known in
the estimation.

- \textit{(ii)} We repeat experiment \textit{(i)} but instead of using the true
observational errors we jointly estimate $\myv Q$ and $\myv R$.

For both experiments, we consider a fully observed state and use the OSS-EnKF
technique.

\begin{figure}[h]
\captionsetup{width=0.5\textwidth}
\centering
\includegraphics[scale=0.5]{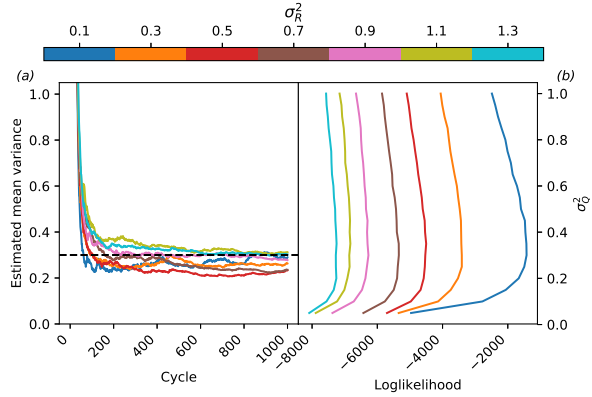}
\caption{(a): Estimated mean model error variance as a function of time for different observational error variances (continuous lines) using the OSS-EnKF implementation. The true value is indicated with a black dashed line. (b): Approximated loglikelihood as a function of the model error variance for different observational error variances (see color bar).}
\label{fig:lor96_q_convergence_true_r}
\end{figure}

Figure \ref{fig:lor96_q_convergence_true_r} (a) shows estimations of the model
error with the Lorenz-96 system for different observational error variances
assuming they are known. This corresponds to experiment \textit{(i)}. Larger
observational errors lead to larger estimations of model error variances and to
a slower convergence in the first cycles. Overall, the technique presents a
reasonable estimation for a rather broad range of observational errors. The
loglikelihood as a function of model error variance for the different
observational error variances is shown in Fig.
\ref{fig:lor96_q_convergence_true_r} (b). The increase of observational error
deteriorates the conditioning of the loglikelihood function, this is in
coherence with \citet{pulido_EM_18}. Furthermore, the maximum of the
loglikelihood function becomes less well-defined for larger values of
$\sigma_R^2$. This can be related to the slower convergence of the estimation.

\begin{figure}[h]
\captionsetup{width=0.5\textwidth}
\centering
\includegraphics[scale=0.5]{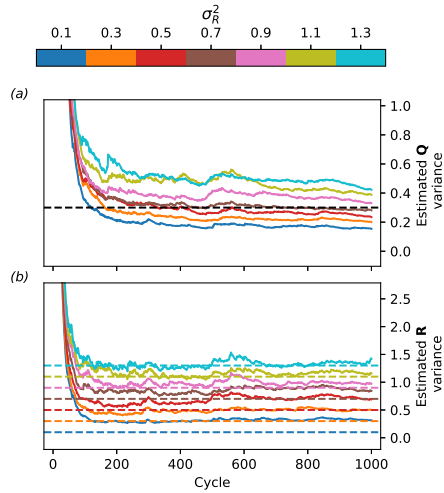}
\caption{Joint estimates of $\myv Q$ (top) and $\myv R$ (bottom) using the OSS-EnKF
implementation for experiments with different true observational error
variances. The dashed lines indicate the true values.}
\label{fig:joint_estimation_convergence}
\end{figure}

For experiment \textit{(ii)} (joint $\myv Q$ and $\myv R$ estimation), Figs. \ref{fig:joint_estimation_convergence} (a) and (b) show the mean estimated
variance of $\myv Q$ and $\myv R$ respectively as a function of time.  The EM
algorithm is able to estimate them jointly and convergence is achieved after
about 400 cycles. The quality of the estimation of $\myv Q$, significantly
degrades in comparison with  experiment \textit{(i)} which uses the true
observational error. The different observational variances are captured rather
well but smaller observational error variances are slightly overestimated. The
systematic error in the estimation of $\myv Q$ increases significantly in
comparison with the experiment with known observational error. For large
observational error variance, the estimations become noisier and the model
error variance is overestimated. In the cases that observational error variance
is overestimated model error variance is underestimated, so that some
compensation between them can be identified.

\begin{figure}[h]
\captionsetup{width=0.5\textwidth}
\centering
\includegraphics[scale=0.6]{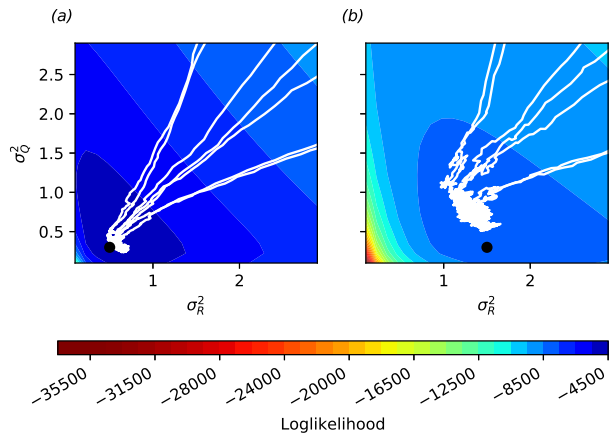}
\caption{The trajectories of the mean of the diagonal of the estimations for
experiments with different initial guess values are shown with white lines. The
total loglikelihood function is represented with the colored contours. The true
value of the parameters is indicated by the black dot. Panel (a) corresponds to
the well-conditioned experiment and (b) to the ill-conditioned one.}
\label{fig:joint_estimation_contours_good_cond}
\end{figure}

As was found in experiment \textit{(i)}, the increase of uncertainty in the
estimation is expected to be associated with a worse conditioning of the
loglikelihood function. Figure \ref{fig:joint_estimation_contours_good_cond}
shows the contours of the loglikelihood as a function of $\sigma_Q^2$ and
$\sigma_R^2$ for observations generated with two different (true) observational
error variances: one to induce a well-conditioned loglikelihood (true values set
as $\sigma_Q^2=0.3$ and $\sigma_R^2=0.5$) and the other to induce an
ill-conditioned loglikelihood function ($\sigma_Q^2=0.3$ and $\sigma_R^2=1.5$).
It also shows the means of the diagonals of the estimations of $\myv Q$ and $\myv R$
for different repetitions of the experiments changing the initial guess
parameters. The likelihood function of the well-conditioned experiment has a
better defined maximum and the estimations are close to the true value, with a
slight tendency to underestimate $\sigma_Q^2$. On the other hand, the
loglikelihood function of the ill-conditioned experiment has a larger
uncertainty in the estimated parameters and there is tendency to overestimate
$\sigma_Q^2$ and underestimate $\sigma_R^2$.

\subsection{Imperfect model experiment}
The experiments described so far have a perfect model structure, the
observations are generated with the same model equations as the ones used to
produce the forecasts in the filter. The developed methodologies are further
assessed in experiments that use an imperfect model.  For these experiments,
we use a two-scale Lorenz-96 model to generate the observations for which only
the large-scale variables are observed. The filter uses a truncated model, a
one-scale Lorenz-96 model, which can only represent the
large-scale variables to produce the forecasts. In this more realistic
situation, the model error is not necessarily Gaussian and indeed it may have
a systematic component. Since we assume the
solution from the complete model is not available, model error comes from using
a truncated model that ignores the small-scale variables. In this scenario we
cannot assess the performance of the proposed methology by comparing the
estimations against the true value, as we did in previous experiments. Hence,
the quality of the estimations is evaluated with respect to an error covariance
matrix which is inferred by comparing the complete and the truncated model
forcing terms which serves as a reference to what the model error should represent.

In the truncated model used to perform the estimation, the forcing term of the
equations for the variable $X_n$ is $F$ while for the true model, it is
$F - \frac{hc}{b} \sum_{j=N_S / N (n-1)+1}^{nN_S / N} Y_j$
(See Appendix \ref{app-A}). Computing the covariance of the difference between
these forcings, assuming a realization \textit{at each model time step}, we get
a covariance matrix structure which serves as a reference to what the model
error should represent. Since the online estimations correspond to the overall
model error covariance within an assimilation window while this
differences correspond to a model time step, a grid search for its best
performing scalar multiple was conducted. The performance of this reference
matrix in the grid search was evaluated considering each multiple as the model
error covariance and computing RMSE and loglikelihood obtained by filtering with
an EnKF. The resulting matrix was then used as a reference to compare to the
estimations. We take $N=8$ large-scale variables and $N_S=256$ small-scale
variables. The small-scale variables are completely unobserved and observations
of the full large-scale system are taken every 50 model time steps.
Observational error covariance is given by $\myv R = 0.5 \myv I_{N_x}$ as
in previous experiments.

\label{sec:results_2scales}
\begin{figure}[h]
\captionsetup{width=0.5\textwidth}
\centering
\includegraphics[scale=0.5]{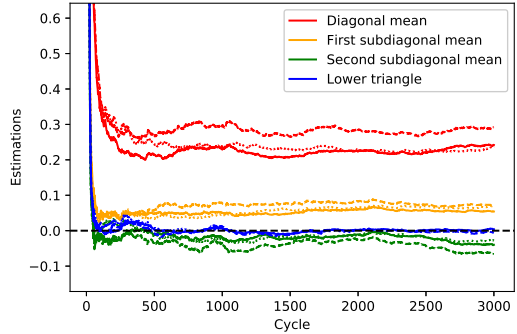}
\caption{Mean of the estimations of the diagonal, first subdiagonal, second sub
diagonal and of the remaining lower triangular matrix for the experiments with
a truncated Lorenz-96 model using the OSS-EnKF (solid line), IS-EnKF(dotted
line) and IS-VMPF(dashed line) implementations.}
\label{fig:2scales_convergence_results}
\end{figure}

Figure \ref{fig:2scales_convergence_results} shows the estimated model error
covariances as a function of time in the imperfect model experiment. The
structure of the estimated covariance gives large values on the diagonal,
smaller values in the first subdiagonal, negative values in the second
subdiagonal and near-zero elsewhere. Overall, there is a rather good agreement
between the estimation of the three implementations, IS-EnKF, IS-VMPF, OSS-EnKF.
The larger difference in the estimated values between the implementations is
found in the variance. The estimated model error variance of this experiment
is between 0.2 - 0.3 which is coherent with previous estimations
(e.g. \citealp{pulido17, scheffler19}).

\begin{figure}[h]
\captionsetup{width=0.5\textwidth}
\centering
\includegraphics[scale=0.6]{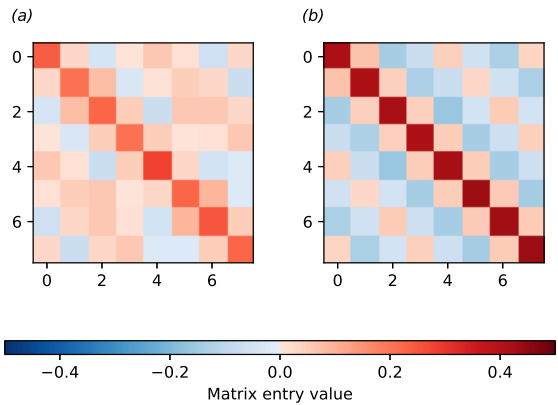}
\caption{(a): Mean of the last 500 estimates of the model error covariance
matrix. (b): Expected covariance matrix obtained from the difference between the
exact and the truncated models of the forcing terms.}
\label{fig:2scales_lor96_heatmaps}
\end{figure}

Figure \ref{fig:2scales_lor96_heatmaps} (a) shows the full structure of the
estimated model error covariance matrix taking the mean of the last 500 OSS-EnKF
estimations in the imperfect model experiment. Figure
\ref{fig:2scales_lor96_heatmaps} (b) shows the reference matrix obtained
offline. The structure of the matrices is similar although entry values of the
estimation are lower than the reference. However, note that the offline
calculation gives the model error covariance at a single model time step while
the estimated one represents the mean model error covariance structure in a
whole assimilation cycle.

%

\section{Conclusions}\label{sec:conclusions}

With the goal of developing an effective estimation technique of error
covariances based on maximization of the likelihood function for nonlinear
hidden Markov models, we proposed an EM-based online scheme that can be coupled
with sequential Monte Carlo filtering techniques and ensemble Kalman filters.
The main objective was to avoid the use of forward-filtering backward-smoothing
strategies since they require to process a batch of observations several times
with high storage and computational costs. While the batch EM requires about
30-100 EM iterations (e.g. \citealp{pulido_EM_18}), each of them imply the
application of the smoother for the whole batch of observations, the proposed
technique estimates model error covariances on the fly together with the
filtering process so that each observation is processed a single time.

The proposed technique shows promising results and is versatile enough to be
applied to different data assimilation scenarios while being computationally
cheaper than the batch EM. The algorithm can accurately estimate the model
error covariance matrix for relatively low-dimensional models with simple error
covariance design. Covariance structures with correlation between neighbouring
variables were recovered. In the case of model error originated by the use of
imperfect model dynamics, the algorithm provided estimates of the covariance
matrix with a shape consistent with the problem. Joint estimation of both model
and observational error is also feasible. Furthermore, the method can track slow
changes in time of the error covariances as shown in the experiments.

A follow-up work will examine the performance in high-dimensional geophysical
systems, for which an apriori parameterization of the covariance matrix is
required. The computational cost is expected to be feasible: apart from
filtering it includes the importance sampling step, the updates of the summary
statistics and the parameters. In high-dimensional systems, the scaling impact
of the sampling noise in the model error estimates needs to be evaluated. It is
expected that localization or regularization through a matrix parameterization
are necessary to constrain sampling noise.

The performance varies with the conditioning of the problem, showing
overestimation of the parameter when the difference between the scale of model
error and observational error is large. Furthermore, the estimates are sensitive
to the choice of the step-sizes, $\gamma_k$. We think that performance can be
improved by incorporating adaptative step-sizes. A potential way to implement
this is by reinterpreting the updates of the estimations as a gradient ascent
and applying optimization techniques such as adaptive learning rates, e.g.
ADADELTA \citep{zeiler12}. This would also circumvent the need of manually
tuning the parameter $\gamma_k$.

Estimation of model error covariance is key for gradient flow filters and other
particle filters which use this information to improve the sampling from the
posterior density in high-dimensional state spaces. In particular, the role of
the model error covariance is essential in the VMPF. It establishes the prior
density as Gaussian mixtures and also has an important role defining the
structure of the reproducing kernel Hilbert space in which the particle flow is
produced. These experiments are the first ones with this filter in which the
structure and values for the model error covariance are not assumed apriori but
are determined entirely through the model error covariance estimates. In this
sense, the results show there is an important positive feedback between the
model error estimation technique and the filter estimates.

\section*{Acknowledgments}
This work was supported by ANPCyT grant number PICT2015-02368.

\bibliography{onlineEM}

\appendix

\section{Appendix. Description of the dynamical systems}\label{app-A}

\subsection{Lorenz-63 system}

This three-variable chaotic model is given by the equations

\begin{align*}
\frac{dx}{dt} &= \sigma (x - y), \\
\frac{dy}{dt} &= x (\rho - z) - y, \\
\frac{dz}{dt} &= x y - \beta z.
\end{align*}

We use as parameter values, the standard ones: $\sigma=10$, $\rho=28$, and
$\beta=8/3$. The integration is performed with a 4th-order Runge-Kutta algorithm
and the used timestep is $\delta t = 0.01$.

\subsection{Lorenz-96 system}

This is an N-variables model given by the equations

\begin{align*}
\frac{d X_n}{dt} = X_{n-1} (X_{n+1} - X_{n-2}) - X_n + F
\end{align*}
for $n = 1,...N$.

Periodic boundary conditions
$X_{-1} = X_{N-1}$, $X_{0} = X_{N}$, $X_{1} = X_{N+1}$ are considered. The
forcing is set to $F=8$ which causes the system to behave chaotically. The
integration is performed with a 4th-order Runge-Kutta algorithm and the timestep
taken to be $\delta t = 0.001$. This time step is chosen small in order to
coincide with the two-scales Lorenz-96 system which needs a finer temporal
resolution for stable numerical integration.

\subsection{Two-scale Lorenz-96 system}

This model considers $N$ large-scale variables given by the equations:
\begin{align*}
\frac{d X_n}{dt} = X_{n-1} (X_{n+1} - X_{n-2}) - X_n + F -
\frac{hc}{b} \sum_{j=N_S / N (n-1)+1}^{nN_S / N} Y_j
\end{align*}
for $n = 1,...N$ and $M$ small-scale variables defined by
\begin{align*}
\frac{d Y_m}{dt} = cb Y_{m+1} (Y_{m-1} - Y_{m+2}) - c Y_m +
\frac{hc}{b} X_{int[(m-1)/N_S/N]+1}
\end{align*}
for $m = 1,...N_S$.
We assume periodic conditions $X_j = X_{j \% N}$ and $Y_j = X_{j \% {N_S}}$ and
set the parameters to $F = 20$, $h=1$, $b=10$, $c = 10$. The integration is
performed with a 4th-order Runge-Kutta algorithm and the used timestep is
$\delta t = 0.001$.

\section{Appendix. Filtering and smoothing techniques}\label{app-B}
\subsection{EnKF equations}

We used here a stochastic version of EnKF based on \citet{burgers98}. For each
time $k$, the particles$ \{ \myv x_k^{a(j)} \}_{j=1}^N$ and
$ \{ \myv x_k^{f(j)} \}_{j=1}^N $ of the analysis and forecast distribution
respectively, are determined from the equations:
\begin{align*}
\myv x_k^{f(j)} &= \mathcal{M}(\myv x_{k-1}^{a(j)}) + \gv \eta_k^{(j)} \\
\myv y_k^{f(j)} &= \mathcal{H}(\myv x_{k}^{f(j)}) + \gv \nu_k^{(j)} \\
\myv K_k \,\, &= \hat{\myv P}_k^f \myv H^T [ \myv H \hat{\myv P}_k^f \myv H^T + \myv R_k]^{-1} \\
\myv x_k^{a(j)} &= \myv x_k^{f(j)} + \myv K_k (\myv y_k - \myv y_k^{f(j)}),
\end{align*}
where $\hat{\myv P}_k^f$ stands for the sample covariance of the forecast
particles $ \{ \myv x_k^{f(j)} \}_{j=1}^N $ and
$\gv \eta_k^{(j)} \sim \mathcal{N}(\myv 0,\myv Q_k)$, $\gv \nu_k^{(j)} \sim \mathcal{N}(\myv 0,\myv R)$.

\subsection{VMPF equations}
Here we briefly describe the VMPF but refer the reader to
\cite{pulido_kernel_18} for a thorough description of it. The VMPF ``moves''
the particles representing the forecast $p(\myv x_{k+1}|\myv y_{1:k})$, towards the
posterior filtering density$p(\myv x_{k+1}|\myv y_{1:k+1})$, through a gradient flow
using a series of small transformations in the directions of the gradient
descent of the Kullback-Liebler divergence between the forecast and the
posterior densities. We use an index $i$ to indicate each of these
transformations, and since we start with the forecast density we associate it
with $i=0$. The algorithm to implement this technique can be sumarized as
follows:

Given a particle representation of the posterior density at time $k$,
$\{ \myv x_k^{(j)} \}_{j=1}^{N_p}$, we evolve the particles with the model to
obtain a representation of the forecast at time $k+1$,
\begin{align*}
 \myv x_{k+1, 0}^{(j)}  = \mathcal{M}( \myv x_k^{(j)}) + \gv \eta_{k+1}.
\end{align*}

Then, a sequence of mappings are applied: each particle is moved following
\begin{align*} \myv x_{k+1, i+1}^{(j)} = \myv x_{k+1, i}^{(j)} -
 \epsilon \nabla \mathcal D_{KL}( \myv x_{k+1, i}^{(j)})
\end{align*}
until a stopping criterion is met. The evaluation of the gradient of the
Kullback-Liebler divergence, $\nabla \mathcal D_{KL}$, in each mapping iteration
can be approximated by Monte Carlo integration by considering the
transformations lie in a reproducing kernel Hilbert space (details in
\citealp{pulido_kernel_18}).

\end{document}